\documentstyle[12pt]{article}

\def\inbar{\,\vrule height1.5ex width.4pt depth0pt}
\def\I1{\relax{\rm 1\kern-.33em 1}}
\def\IK{\relax{\rm I\kern-.18em K}}
\def\IQ{\relax\hbox{$\inbar\kern-.3em{\rm Q}$}}
\font\cmss=cmss10 \font\cmsss=cmss10 at 7pt
\def\IZ{\relax\ifmmode\mathchoice
{\hbox{\cmss Z\kern-.4em Z}}{\hbox{\cmss Z\kern-.4em Z}}
{\lower.9pt\hbox{\cmsss Z\kern-.4em Z}}
{\lower1.2pt\hbox{\cmsss Z\kern-.4em Z}}\else{\cmss Z\kern-.4em Z}\fi}

\let\a=\alpha
\let\b=\beta

\let\d=\delta
\let\e=\epsilon

\let\th=\theta

\let\l=\lambda
\let\m=\mu

\let\r=\rho

\let\vf=\varphi

\let\o=\omega

\let\D=\Delta

\def\ovr{\over}

\def\beq{\begin{equation}}
\def\eeq{\end{equation}}
\def\bear{\begin{eqnarray}}
\def\enar{\end{eqnarray}}
\def\beit{\begin{itemize}}
\def\enit{\end{itemize}}
\def\ld#1{\partial_{#1}}

\def\+-{\underline +}
\def\1/2{{1\over 2}}
\def\nonum{\nonumber \\}
\def\ln{{\rm ln}}

\def\tr{{\,\rm Tr\,}}
\def\zb{\bar{z}}

\def\cE{{\cal E}}
\def\ob{\bar \o}

\def\I1{\relax{\rm 1\kern-.33em 1}}
\def\br{{\bf r}}

\def\tr{{\,\rm Tr\,}}

\def\ln{{\rm ln}}
\def\zb{{\bar z}}
\def\cG{{\cal G}}
\def\diag{{\rm diag}}

\def\pd{{\,\partial\,}}
\def\Re{{\rm Re}}

\newcommand{\sect}[1]{\section{#1} \setcounter{equation}{0}}
\renewcommand{\theequation}{\thesection.\arabic{equation}}

\begin{document}
\titlepage
\begin{flushright}
QMW-94/29 \\
August 1994\\
\end{flushright}
\begin{center}
	{\bf\Large Screening in Two Dimensional Nonabelian Vortex Systems}\\
\rm
\vspace{3ex}
			  CHINORAT KOBDAJ
	\footnote{Work supported by Thai Government.
	\\ \indent \hskip .5cm e-mail: C.Kobdaj@qmw.ac.uk}
\vspace{1ex}
				\\and\\
			    STEVEN THOMAS
		\footnote{ e-mail: S.Thomas@qmw.ac.uk}\\
\vspace{2ex}
		{\it Department of Physics \\
		Queen Mary and Westfield College \\
			Mile End Road\\
			  London E1 4NS\\
			    U.K.}\\
\vspace{2ex}
			  ABSTRACT\\
\end{center}
\noindent
We study charge screening in a system of two dimensional nonabelian
vortices, at finite temperature. Such vortices are generated after an \(
SO(3) \) global symmetry group is spontaneously broken to a discrete
subgroup
\( \IQ_8 \), where the latter is isomorphic to the nonabelian group of
quaternions. Poisson-Boltzmann like equations are derived for the various
inter-vortex potentials, and the solutions to lowest order in a small
fugacity expansion, are shown to behave much like those in abelian or
Coulomb charge systems.  The consequences for the phase structure of the
system are discussed, where the fugacities associated to the thermal
production of the various types of nonabelian vortices, are shown to play a
key role.

\newpage
\sect{Introduction}
 Vortex defects in two and three spatial dimensions, have many interesting
 properties which have found a wide variety of applications in two notable
 areas, condensed matter physics [1], and cosmology [2]. Indeed it is well
 known [3] that the occurrence of vortices in 3-dimensional condensed
 systems such as superfluid helium IV, play an important role in
 understanding the physical properties of these systems.

Vortex defects in any dimension are characterized by \( \pi_1 (M ) \), the
fundamental group of the vacuum manifold \( M \) of the theory.  The type
of vortex defects that occur in for example helium IV are `abelian' in the
sense that they are characterized by an abelian fundamental group,
isomorphic to the group of integers \( \IZ \).  In 2-spatial dimensions,
abelian vortices have particularly simple interactions which allows one to
write down the grand canonical partition function for thermal pair creation
to all orders in the vortex fugacity in terms of the so called Coulomb gas
model [4].  The pioneering work of Berezinsky [5], and Kosterlitz and
Thouless [6], into their statistical mechanical properties, showed that a
gas of such vortices underwent a novel kind of phase transition at some
critical temperature \( {\rm T}_{\rm c} \). A simple physical picture
emerged where for T \( < {\rm T}_{\rm c} \), vortices and antivortices form
a medium of bound pairs which subsequently dissociates into free vortices
and antivortices for T \( >{\rm T}_{\rm c} \).  These results when applied
to approximately 2-dimensional systems such as helium IV thin films, lead
Kosterlitz and Nelson [7] to predict a universal jump in the superfluid
density at \( {\rm T}_{\rm c} \), which was later experimentally verified
[8].

In a previous publication [9], we presented the results of some preliminary
investigations into the nature of interactions of nonabelian vortices in a
particular 2-dimensional model.  In this paper we shall focus on the
problem of screening of the forces between nonabelian vortices, again in
the context of the simple model considered in [9].  Ordinary abelian charge
screening is the mechanism behind the Kosterlitz-Thouless phase transition
[1].  Although there are a number of ways of exhibiting this screening, for
example by exploiting the map between the vortex gas partition function and
that of sine-Gordon theory in 2-dimensions [10], perhaps the most
illuminating is to derive a so called Poisson-Boltzmann (P-B) equation
[11], satisfied by the linearly screened potential between a test vortex
and antivortex placed in the gas.  One of the goals of the present paper is
to derive a P-B equation for the various inter-vortex potentials for the
case of nonabelian vortices.  In the model under consideration, there are
basically 2 species or types of vortices in the nonabelian system at finite
temperature.  We shall show that to first order in a small fugacity
expansion, the P-B equations effectively become those of a pair of
`abelian' P-B equations , one for each vortex type. However because of the
nonabelian nature of the vortices in question, the system is not simply
reducible into two non-interacting `Coulomb' systems to this order. In
fact, because of the nonabelian fundamental group \( \pi_1 (M ) \) , we
find that the coupling constants of the two Coulomb systems are
constrained.

The structure of the paper is as follows. In Section 2 we give a brief
review of the 2-dimensional field theory model [12] which describes
nonabelian vortices associated with \( \pi_1 ( M ) \cong \IQ_8 \), and
which was studied further in [9].  Vortices corresponding to this
fundamental group may have some relevance to certain liquid crystals [13].
In section 3, as an introduction to Poisson-Boltzmann equations and to fix
notation, we show how to derive the P-B equation for a Coulomb gas, which
is known to be physically equivalent to a system of abelian vortices.
Sections 4 and 5 are concerned with deriving the same for the nonabelian
vortex system defined in section2, and some conclusions are drawn about the
possible phase structure of the system.

\sect{A model of two dimensional  nonabelian vortices}
Perhaps the simplest model [12] in which nonabelian vortex defects occur,
is that in which spontaneous symmetry breaking occurs in two spatial
dimensions, with an order parameter \( \Phi \) characterizing a system
whose total energy \( E \) we may choose to be
\begin{equation}\label{eq:2.1}
    E=\int d^2\, x \
	   \1/2 \tr\left[ g^{ab} \partial_a \Phi
			\partial_b \Phi\right] + V(\Phi)
\end{equation}
\bear\label{eq:2.2}
	V(\Phi) & =&
		{\l\ovr 4}\tr \Phi^4+{\l'\ovr 4}(\tr\Phi^2)^2\nonum
&&      + {\r\ovr 3}\tr\Phi^3-\1/2\m_0^2\tr\Phi^2
\enar
where \(\Phi\) is a scalar field transforming in the five dimensional
representation of the symmetry group \( G= SO(3)\), i.e. it is a traceless
\( 3 \times 3 \) symmetric matrix.

When \(\r\not=0\), \(V \) has three isolated minima and the unbroken
symmetry group is \( U(1)\). This case is not particularly interesting
since there are no stable vortex defects produced.  This is a result of a
topological theorem which states that if the first homotopy group \(\pi_1(
G/H)\) is trivial, where \( H\) is the unbroken symmetry group, then the
vortices are unstable, and indeed in this case
\(\pi_1( SO(3)/U(1))\simeq \I1.\)
If \(\r=0\) however, \( SO(3)\) is broken down to a discrete subgroup
\(\IK\) isomorphic to \(\IZ_4\) the additive group of order 4, and in this
case there is a degenerate set of minima of \( V\) which lie on the ellipse
\( \vf_1 ^2 + \vf_2^2 + \vf_1 \vf_2 \,
  = \, {{\mu_{0}}^2 }/{( \lambda + \lambda^{'} )}\)
, where the latter two fields are the independent eigenvalues of the
diagonalized field \( \Phi \).  The fundamental group of the manifold of
vacuum states of the theory defined in eqs.(\ref{eq:2.1}) and
(\ref{eq:2.2}), is given by
\beq\label{eq:2.3}
 \pi_1(SO(3)/ \IK)\simeq \IQ_8
 \eeq
\( \IQ_8 \) being  a nonabelian discrete group of order 8,
(isomorphic to the group of quaternions). It is generated by elements \(i,
j, k \), \(-i, -j, -k\) and -\I1, with
\bear\label{eq:2.4}
 {i}^2 = {j}^2 = { k}^2 = -\I1 \nonum i\, j \,= \, k \quad ,\quad  j \,
k \, = \, i \quad,\quad k\, i\, = \,j
\enar
where \( i,j,k\) define a basis of quaternions. Hence there are three types
of nonabelian vortices in this model corresponding to \(i,j\) and
\(k\), ( the elements \(-i , -j \) and \( -k \) refer to the corresponding
anti-vortices, whilst \( -\I1 \) actually defines a \(\IZ_2 \) vortex which
is abelian in nature since \(-\I1 \) commutes with all the other elements
in \( \IQ_8 \) ).  The vortices which are produced by this symmetry
breaking are guaranteed to be stable by topological arguments.

In ref.[9], the following ansatz was used to minimize (numerically)
the energy \( E \)
\beq\label{eq:2.5}
	\Phi(r,\theta) \ = \ {\cal G }(\theta)\, \Phi_{\rm diag}(r) \,
				\,{\cal G}^{-1}(\theta),
\eeq
where the diagonal matrix
\(\Phi_{\rm diag}=(\vf_1,\vf_2,-(\vf_1+\vf_2))\). The \( SO(3) \)
group elements
\( {\cal G}(\theta) \) satisfy the nonabelian vortex boundary conditions
\beq\label{eq:2.6}
      {\cal G}(\theta + 2\,\pi )\, = \,{\cal G}(\theta )\, h
\eeq
where \(h\) is an element of the group \(\IK\), and (\(r,\theta\)) are
 polar coordinates centered on the vortex core. Explicit solutions for the
 functions \(\vf_ 1,\vf_2 \) and \({\cal G} \) pertinent to describing
 vortices of various types, were given in [9]. Moreover, it was also shown
 that the ansatz in eq.(\ref{eq:2.5}) was stable to perturbations that
 preserve the boundary conditions, eq.(\ref{eq:2.6}). In this sense, lowest
 energy vortices are correctly described by eq.(\ref{eq:2.5}).

For the energy of the vortex to be finite, \(\Phi(r,\theta)\) must tend
asymptotically as \(r\to\infty\), to a minima of the potential
\(V(\Phi)\). In addition, \(\Phi(r,\theta)\) should vanish sufficiently
quickly as \(r\to 0\). Classical field  configurations \(\Phi\) with
these properties were obtained by numerical minimization of the
energy \(E\) and the reader is referred  to  [9] for further details.

Having mentioned the stability of vortices described by eq.(\ref{eq:2.5})
to perturbations that maintain the homotopy classes, one still has to be
cautious before concluding that such vortices will be stable if produced
thermally in a system at finite temperature. This is because dissociation
can occur between one vortex type and another, depending on the relative
size of the chemical potentials involved. From a group theoretic point of
view, this process of dissociation can be understood simply as a
consequence of group multiplication, e.g.  the relation \( i j = k \) in
eq.(\ref{eq:2.4}) implies that a \( k \) type vortices could decay into an
\( i \) and \( j \) type.  In [9] it was shown that the chemical
potentials controlling the concentration of the various types of vortices,
depends on the single real parameter \( a = {\displaystyle \frac{< \vf_1
>}{<\vf_2 >}} \),
which essentially fixes a point on the ellipse of vacuum states in this
model. As the value of this parameter is varied, either one of \( i, j \)
or \( k \) type vortices are unstable to dissociation into the other
two. For definiteness, we will restrict \( a \) to be in a range such that
\( k \) types are unstable to dissociation into those of \( i \) and \( j
\). (In passing, it was proved in [9] that the abelian vortices
corresponding to the element \( -\I1 \) always dissociates into
vortex-antivortex pairs for any choice of \( a\), so we will ignore these
in the remain of the paper). To summarize, kinematic constraints simplify
the study of nonabelian vortices at finite temperature, at least in the
model under consideration, by allowing us to focus on just two types of
vortices corresponding to particular non-commuting elements of \( \IQ_8 \).

It is the purpose of this paper to try and understand the screening
properties of the (\( i \) and \( j \) ) type nonabelian vortices, by
deriving Poisson-Boltzmann (P-B) like equations for the corresponding
inter-vortex potentials. Since the P-B equations are more familiar in the
context of screening in systems of abelian vortices, in the next section we
shall review their derivation, and the consequences they have for
Kosterlitz-Thouless type phase transitions.

\sect{Coulomb gas system}
It is well known that a system of abelian vortices in 2-dimensions can be
mapped onto that of the so called Coulomb gas [4]. Abelian means that the
fundamental group associated with spontaneous symmetry breaking is itself
abelian. Such vortices occur when \( U(1) \) symmetry is broken e.g.  in
superfluids, superconductors etc. [1]. Because of the mapping between
systems of Coulomb charges and abelian vortices, the notions of charge and
vorticity are interchangeable, and we shall often use both sets of
terminology in what follows.  Consider then, a Coulomb system of N charged
particles, with charges \( \pm q \), and let \(N_+ \) ( \( N_-
\) ) be the total number of positive (negative ) charges with
\( N = N_+ + N_- \). Moreover, let us restrict ourselves to neutral Coulomb
gases, i.e. those in which \( N_+ = N_- \).  The total energy of such a
configuration is given by
\beq\label{eq:3.1}
	H_N\,=\sum_{a \not =b=1}^N \,\1/2 q_a q_b U(r_{ab})-\,N\,\mu
\eeq
 where in eq.(\ref{eq:3.1}), \(U(r_{ab})\) is the electrostatic energy of
 charge \(q_a\) due to charge \(q_b\) and \(r_{ab}=|\br_a-\br_b|\) is the
 distance between those charges. The chemical potential \(\mu\) is a sum of
 the electrostatic self-energy \(U(0)\) and non-electrostatic contribution,
 \(E_c\) or the core energy.
\beq\label{eq:3.2}
	\mu\,=\, -[\1/2 U(0) +E_c]
\eeq
The partition function of the system is
\beq\label{eq:3.3}
	Z\,=\,\sum_{N\, =\,0}^\infty\frac{1}{N!}
	\sum_{\{ q_a;  {\displaystyle \,\Sigma} q_a \, = 0 \}}	\prod_{a=1}^N
	\int \frac{d^2 \br_a}{ \pi\,\xi^2}  e^{\,- H_{N}/T}
\eeq
  where \(\pi \xi^2 \) is a phase-space division factor and \( \xi \) is a
 length scale typically the size of a Coulomb particle.  In
 eq.(\ref{eq:3.3}) and subsequently, we set the Boltzmann constant \( k_B
 \, = \, 1 \).  For a system of interacting abelian vortices, \( \xi \)
 corresponds to the core radius.  The restriction \(\sum q_a=\,0\) on the
 sum over charges in eq.(\ref{eq:3.3}) follows from the assumption that we
 are considering a neutral system. Introducing two infinitesimal test
 charges \(\delta q_+,\,\d q_-\) into the system at fixed positions
 \(\br_+\) and \(\br_-\), we can approximate the partition function of this
 modified system as follows
\bear\label{eq:3.4}
 \lefteqn{Z(\d q ,\br_+ , \br_-) \  =\ 1+\int  d^2 \br  \ e^{-
				\1/2\d q_+ \d q_- V_L(r)/T }
              \d^2(\br-(\br_+-\br_-))      }   \nonum
   && \qquad +  \int \frac{d^2 \br}{\pi\,\xi^2}\  e^{-\left[ {1\over 2}
			\d q_+  q_+ V_L(\br-\br_+) + {1 \over 2} V_L(0)
		         +E_c
			\right]/T
                  -\left[ {1 \over 2}
			\d q_-  q_+ V_L(\br-\br_-) +{1 \over 2} V_L(0)
		         +E_c
		\right]/T}                    \nonum
   && \qquad +  \int \frac{d^2 \br}{\pi\,\xi^2}\  e^{-\left[ {1\over 2}
			\d q_+  q_- V_L(\br-\br_+) +{1 \over 2}  V_L(0)
		         +E_c
			\right]/T
                   -\left[ {1 \over 2}
			\d q_-  q_- V_L(\br-\br_-) +{1 \over 2} V_L(0)
		         +E_c
  		\right]/T}
	   \nonum
   && \qquad +\quad ...
\enar
where in eq.(\ref{eq:3.4}), we have only kept terms at most linear in the
fugacity \( z \, = \, {e}^{-E_c /T} \), which is taken to be small.

 Since the two test charges introduced into the Coulomb system are not
thermally created, there will be no chemical potentials appearing in the
second term of eq.(\ref{eq:3.4}).  {\mbox{\(V_L(\br) \)} } is the linearly
screened potential at \(\br \) which in the absence of background charges,
is simply the unscreened Coulomb potential depending logarithmically on
distance between charges.  The interaction between the test charges \(\d
q_+\) and \(\d q_-\) and the background charges \(q_+\) and \(q_-\) which
are thermally created, depends on the core energy \(E_c\) and the
self-energy \(V_L(0)\).  We define \(\d q_+ = -\d q_-\,=-\d q \), \(
\,q_+=-\,q_- =\,q\) and the density of positive(negative) background
Coulomb gas charges \(n_F^{(\pm)} \) as
\beq\label{eq:3.5}
	n_F^{(\pm)}\,=\,\frac{e^{(-{1\over 2} V_L(0) -E_c)/T}}{\pi\,\xi^2}\,
	      =\, \frac{z}{\pi\,\xi^2}\,e^{-  V_L(0)/2T}
\eeq
then we may rewrite the partition function as
\bear\label{eq:3.6}
	&Z(\d q ,\br_+ , \br_- )= &1+\int  d^2 \br  \
		e^{\1/2(\d q)^2 V_L(\br)/T } \d^2 (\br-(\br_+-\br_-))
              \nonum
	   && \quad +   \,n_F^+
			\int \frac{d^2 \br}{\pi\,\xi^2}\  e^{
			-{1\over 2}\d q\,  q\, V_L(\br-\br_+)/T
			\ +{1 \over 2}\d q\,  q\, V_L(\br-\br_-)/T}
              \nonum
	   && \quad +  \,n_F^-
			\int \frac{d^2 \br}{\pi\,\xi^2} \ e^{
			+{1 \over 2}\d q\,  q\,V_L(\br-\br_+)/T
			\ -{1 \over 2}\d q\,  q\, V_L(\br-\br_-)/T}
			 +\,...             \nonum
\enar
 To obtain the linear screening potential at the point \( \br_{0} =
(\br_+-\br_-\)), we apply techniques which are familiar in field
theory. First introduce a source term or test charge into the generating
function or partition function (as we have done above). Then,
differentiating the new partition function with respect to these sources
gives us a correlation function which is proportional to \(V_L\),
\beq\label{eq:3.7}
\d q V_L(\br_+-\br_-)\,
   =\, T\, \frac{\partial \, \ln\,Z(\d q,\br_+ , \br_- )}{\partial \,\d q}
   =\,{T \ovr Z(\d q ,\br_+ , \br_-)}
	\frac{\partial Z(\d q ,\br_+ , \br_- )}{\partial \,\d q}
\eeq
Thus,
\bear\label{eq:3.8}
  \lefteqn{ \d q V_L(\br_+-\br_-) \
	     =\ {\displaystyle {T\ovr Z}}\Biggl\{\,\d q \,V_L(\br_+-\br_-)\,
		 e  ^{\1/2 (\d q)^2 V_L(\br_+-\br_-)/T } }
		\nonum
	    &  +       \,{\displaystyle {\frac{q\,n_F^+}{2\,T}}
	             \int} d^2 \br \ [- V_L(\br-\br_+)+  V_L(\br-\br_-)]
			e^{       -{1\over2}\d q\,  q\, V_L(\br-\br_+)/T
				\ +{1\over2}\d q\,  q\, V_L(\br-\br_-)/T}
              \nonum
	    & \quad +\,{ \displaystyle {\frac{q\,n_F^-}{2\,T}}
		     \int }d^2 \br \ [ V_L(\br-\br_+)-  V_L(\br-\br_-)]
			e^{        {1\over2}\d q\,  q\, V_L(\br-\br_+)/T
				\ -{1\over2}\d q\,  q\, V_L(\br-\br_-)/T}
		 + ... \Biggr\} \nonum
\enar
 Since we assume small fugacity \(z \) and shall ignore nonlinear terms in
\(\d q\) (by taking \(\d q\) sufficiently small), one can approximate
\(1/Z \) by unity (\( Z\,=\,1+O(z)\)). To obtain the Poisson-Boltzmann
equation [1, 11], we apply the two dimensional Laplacian \(\nabla^2_{r_0} \)
with respect to the relative position of the test charges \({\bf
r}_0=({\br_+-\br_-})\) to both sides of eq.(\ref{eq:3.8})
\bear\label{eq:3.9}
\lefteqn{\d q \nabla^2_{ {\bf r}_0} V_L(\br_+-\br_-)
 \ \approx \ \d q\, \nabla^2_{ {\bf r}_0} V_0(\br_+-\br_-)
	e ^{\1/2(\d q)^2 V_L(\br_+-\br_-)/T } }
\nonum
& + \,{ \displaystyle {\frac{q\,n_F^+}{2}}
	\int }d^2 \br \ \nabla^2_{ {\bf r}_0}
	[-V_0(\br-\br_+)+ V_0(\br-\br_-)]
	e^{ -{1\over2}\d q\, q\, V_L(\br-\br_+)/T \ +{1\over2}\d q\, q\,
	V_L(\br-\br_-)/T}
\nonum
& \quad
	+\,{ \displaystyle {\frac{q\,n_F^-}{2}}
	\int }d^2 \br \ \nabla^2_{ {\bf r}_0}
	[V_0(\br-\br_+)- V_0(\br-\br_-)]
	e^{{1\over2}\d q\, q\, V_L(\br-\br_+)/T \
	-{1\over2}\d q\, q\,V_L(\br-\br_-)/T} + ...  \nonum
\enar
 where \(V_0(\br)\) is the lowest order (unscreened) approximation of
 \(V_L(\br)\). Since \(V_0(\br)\) is the interaction potential of charge
 particles in two spatial dimensions, it is proportional to the logarithmic
 function, so that
\beq \label{eq:3.10}
	\nabla^2_{\br_0} V_0(\br-\br_+)=-2\pi\,\a \d^2(\br-\br_+)
\eeq
 where \( \a \) is a constant with dimensions of temperature, and is the
 effective coupling constant of the Coulomb gas.  Thus we obtain the
 Poisson-Boltzmann equation as
\bear\label{eq:3.11}
 	\d q \nabla^2 V_L(\br_+-\br_-)
 	&=& \,-  2\pi \a \d q\, \,\d^2 (\br_+-\br_-)
	\nonum
	&&
		+\,\frac{2\pi\a\,q\,n_F}{2}
		 e^{{1 \over 2}\d q\,q\, [V_L(\br_+-\br_-)-V_L(0)]/T }
	\nonum
	&&
		-\,\frac{2\pi\a \,q\,n_F}{2}
		 e^{-{1 \over 2}\d q\,q\, [V_L(\br_+-\br-)-V_L(0)]/T }
\enar
 where \( n_F\,=\, n_F^+ + n_F^- \) is the number density of free charges.
 Next, we redefine \\ \(V_L(\br_+-\br_-)\) as the separation energy between
 a test charge and background charge,
\[
	E_{Sep}=V_L(\br_+-\br_-)-V_L(0) \to {\tilde V}_L(\br_+-\br_-)
\]
where \(V_L(0)\) is the energy of zero separation between test charge and
 background charge. Finally, dropping the \~\ (tilde)\ and expanding the
 exponentials to lowest order in \( \d q \) we obtain the Poisson-Boltzmann
 equation corresponding to an abelian (in the sense defined earlier)
 Coulomb gas
\beq\label{eq:3.12}
	\d\, q \nabla^2 V_L(\br_+-\br_-)\,=\,
	      -   2\pi\a\,\d\, q\,  \d^2(\br_+-\br_-)
              +\,\frac{\pi\a\,\d\, q\,q^2\,n_F}{T}
			 V_L(\br_+-\br_-)
\eeq
Solutions to eq.(\ref{eq:3.12}) have been discussed in the literature
[1,11] in the context of charge unbinding in the Coulomb system. It is
clear that when \( n_F \neq 0 \), the potential \( V_L \) is screened, and
decreases exponentially with the separation between a pair of opposite
charges (or vortex-antivortex pair in the case of abelian
vortices). Consequently, the pairs become unstable to dissociation into
free charges, and so `unbinding' of the charge or vorticity occurs. This in
turn implies of course, a non-vanishing density of free charges \( n_F \)
so is a consistent physical picture. The alternative possibility is that
\mbox{\( n_F = 0 \)}, in which case the P-B equation for \( V_L \) implies
that the latter is logarithmically decreasing with the pair separation,
giving rise to a medium of bound neutral pairs. The question of whether
\(n_F \) is vanishing or not depends on the temperature \( T \) of the
system, i.e. \( n_F = 0 \) if
\( T < T_c \), and \( n_F \neq 0 \) if \( T> T_c \),
where the critical temperature
\( T_c =  {\displaystyle {\a \ovr 4 }}( 1 - 2 \pi z )  \)
define a line of values as \( z \) is varied [5,6] . \( T_c \) marks the
position of the so called Kosterlitz-Thouless phase transition of the
Coulomb gas [6] . Perhaps the simplest way of obtaining the leading order (
in \( z \) ) approximation of \( T_c \), (i.e. \( T_c = {\displaystyle {\a
\ovr 4 }} \) ) is by the method of self-consistent screening [7].  One
introduces the so called screening length \( \omega (z, T) \), which for
an infinite system is related to the density of free charges \( n_F \) by
\beq\label{eq:3.14}
	n_F (z,T) \, = \, {T \ovr {\pi \alpha\omega^2 (z,T)}}
\eeq
where it is clear from eq.(\ref{eq:3.14}) that \( \omega \) depends on the
fugacity \( z \) and \( T \). At the same time, since the linearly screened
potential at the origin
\( V_L(0) \, = \, {\rm ln} ( { \displaystyle \omega\ovr\xi} ) \),
eqs.(\ref{eq:3.5}) and
(\ref{eq:3.14}) together provide a consistent set of equations determining
the value of \( \omega \) . There are two self-consistent solutions namely
for \( T > T_c \), \( \omega\) is finite (and \( V_L \) exponentially
decays ) whilst for \( T<T_c \) the screening length is infinite and the
potential remains logarithmic, with \( T_c \, = { \displaystyle {\a \over
4}}\) [7].  Alternatively, one can deduce the dependence of \( \lambda \)
on \( z \) and \( T\), by utilising the well known map of the Coulomb gas
onto the sine-Gordon model [10], and exploiting the known renormalization
properties of the latter. This method also gives the next to leading order
form of \( T_c \) given above.  In section 5, we shall return to the
problem of solving the Poisson-Boltzmann equation for systems of nonabelian
vortices or `charges', which we shall introduce in the next section.
\sect{System of nonabelian vortices.}
We now wish to turn our attentions to a system of two dimensional
nonabelian vortices of type \( i\) and \(j \), (i.e. field configurations
satisfying the ansatz of eq.(\ref{eq:2.5}) with twisted boundary conditions
as in eq.(\ref{eq:2.6}) with \( h = i \) and \( j\) respectively.)  To
construct the Poisson-Boltzmann equation for the nonabelian vortex system,
we first have to define the infinitesimal charge of a nonabelian
vortex. From our ansatz eq.(\ref{eq:2.5}), the field configuration \(\Phi\)
depends on the \( SO(3) \) group elements \({\cal G}_\a \), which represent
rotations about the \(x,y,\) or \(z\) axes. Therefore, the infinitesimal
charges are defined correspondingly as rotations through an infinitesimal
angle \(\e\,\theta\) about those axes. For example, the classical field
configuration \(\Phi\) of an infinitesimal test charge of type \(i\) is
\beq\label{eq:4.1}
	\Phi_{\e i}= \,{\cal G}_{\e\, i} \Phi_{diag}{\cal G}^{-1}_{\e\, i}
\eeq
where
\beq\label{eq:4.2}
{\cal G}_{\e\,i} =
\pmatrix{ 1 &  0              &              0 \cr
 0 & \cos ({{\displaystyle \e\,\th\ovr 2}} )
   & \sin({{\displaystyle\e\,\th\ovr 2}} ) \cr
 0 & -\sin ({{\displaystyle\e\,\th\ovr 2}} )
   & \cos ({{\displaystyle\e\,\th\ovr 2}} ) \cr
	}
\eeq
with similar definitions for the vortices of type \(j\) and \(k\). One can
compute the infinitesimal nonabelian vorticity or nonabelian charge of
these classical field configurations, by for example calculating the form
of the self energy in each case and comparing it to the abelian case.  One
finds
\bear\label{eq:4.3}
 \d q_{ i} &=\,-\d q_{- i} &= \frac{\e}{\sqrt{2}}(\vf_1+2\,\vf_2) \nonum
 \d q_{ j} &= \,-\d q_{- j}&=\frac{\e}{\sqrt{2}}(2\,\vf_1+\vf_2) \nonum
 \d q_{ k} &= \,-\d q_{- k}&=\frac{\e}{\sqrt{2}} (\,\vf_1-\vf_2) \nonum
\enar
with infinitesimal parameter \(\e\). The corresponding finite nonabelian
charges are
\bear\label{eq:4.4}
 q_{ i}   &= -\,q_{- i}   &=  \frac{1}{\sqrt{2}}(\vf_1+2\,\vf_2) \nonum
 q_{ j}   &= -\,q_{- j}   &=  \frac{1}{\sqrt{2}} (2\,\vf_1+\vf_2) \nonum
 q_{ k}   &= -\,q_{- k}   &=   \frac{1}{\sqrt{2}}( \vf_1-\vf_2)
\enar
It should be noted that the ``charges'' defined in eq.(\ref{eq:4.4}) have
dimensions of \( \sqrt{\rm energy } \), compared to the dimensionless
ones introduced in section 3.

Having discussed how to define the notion of infinitesimal charge for
nonabelian vortices, we are now in a position to consider the `nonabelian'
charge-unbinding scheme via the Poisson-Boltzmann description. Let \( (
N_{-i} \, , N_{-j} ) N_{+i} \, , N_{+j} \) be the total number of \(i\) and
\(j \) type (anti) vortices present. As usual we define \( N_i = N_{+i} +
N_{-i} \) and \( N_j = N_{+j} + N_{-j} \) and again restrict ourselves to
neutral systems i.e.  to configurations having equal numbers of vortices
and antivortices of the same type, so \( N_{+i} \, =\, N_{-i} \) similarly
for \(j\) types.

Under these circumstances, the partition function of the system is
\bear\label{eq:4.5}
	Z\,&=&\,\sum_{N_i=0 }^\infty \sum_{N_j=0 }^\infty
	\frac{1}{N_i!N_j!}
		\prod_{a_i=1}^{N_i}\prod_{a_j=1}^{N_j}
	\sum_{ q_{a_i}	;{\displaystyle \Sigma}_{a_i} q_{a_i} = 0 }\,
 	\sum_{ q_{a_j}	;{\displaystyle \Sigma}_{a_j} q_{a_j} = 0}\,\cr
	 && \times \quad \int \frac{d^2 \br_{a_i}}{ \pi\,\xi^2}
		\int \frac{d^2 \br_{a_j}}{ \pi\,\xi^2}e^{\,- H(N_i,N_j)/T}
\enar
Generally, one can write down the hamiltonian of the nonabelian system as
\bear\label{eq:4.6}
	H(N_i,N_j)\,
	&=&\sum_{a_i \not=b_i=1}^{N_i} \,\1/2 q_{a_i} q_{b_i}
		U_{ii}(r_{a_i} -r_{b_i})
	  +\sum_{a_j \not=b_j=1}^{N_j}\,
		\1/2 q_{a_j} q_{b_j} U_{jj}(r_{a_j} - r_{b_j}) \cr
	&+&
	   \sum_{a_i, \, b_j=1}^{N_i, N_j}\, \1/2 q_{a_i} q_{b_j}
	 U_{ij} (r_{a_i} - r_{b_j} ) -\mu_i N_i - \mu_j N_j
\enar
 The quantity \(U_{ii}(r_{a_i}-r_{b_i})\) is the interaction energy due to
the charges \( q_{a_i} \, , q_{b_i} \) of the \( i\) type vortices amongst
themselves, with similar definitions for \( U_{jj} \) and \( U_{ij} \). The
chemical potentials \(\mu_i\) and \(\mu_j\) are
\bear\label{eq:4.7}
	\mu\,&=&\, -[\1/2 U_{ii}(0)+E_c^i] \nonum
	\mu\,&=&\, -[\1/2 U_{jj}(0)+E_c^j]
\enar
where \( E^{i}_c \) and \( E^{j}_c \) are the corresponding core energies
of \( i \) and \( j \) type vortices respectively, explicit expression for
which may be found in [9].  We define the fugacities corresponding to the
thermal creation of vortices of type \( i\) and \(j\) respectively
\beq\label{eq:4.8}
	z_i=e^{-E_c^i/T} ,\quad
	z_j=e^{-E_c^j/T}.
\eeq
Now we introduce two infinitesimal test charges \(\d q_{ i}\) and \(\d q_{-
i}\) into our system at the points \(\br_{\e i}\) and \( \br_{-\e i} \)
where we remind the reader that only the vortices of type \(i\) and \(j\)
need be considered if we restrict the parameter \( a \) in the range
(\(-2<a<-\1/2\)) ).

Then, expanding the partition function denoted by \mbox{ \(Z_{i}(\d
q_i,\br_{\e i}, \br_{-\e i} ) \)} for the modified system in the presence
of such test charges to linear order in the fugacities (which are again
taken to be small), we obtain
\bear\label{eq:4.9}
 &Z_{i}( \d  q_i, &\br_{\e i},\br_{-\e i} ) \,
		= 1+\int  d^2 \br  \ e^{- \1/2\d q_{ i} \d q_{- i}
		V^{i i}_L(\br)/T } \d^2 (\br-(\br_{\e i}-\br_{-\e i}))
             \nonum
	   \quad &+&z_i              \int d^2 \br\
		e^{-[E_{\e ii}(r-r_{\e i})+E_{-\e ii}(r-r_{-\e i})]/T }
              \nonum
	    \quad &+&z_i              \int d^2 \br \
		e^{-[E_{\e i-i}(r-r_{\e i})+ E_{-\e i-i}(r-r_{-\e i})]/T }
	      \nonum
	  \quad  &+&z_j              \int d^2 \br\
		e^{-[E_{\e ij}(r-r_{\e i})+E_{-\e ij}(r-r_{-\e i})]/T }
              \nonum
	   \quad &+& z_j              \int d^2 \br \
		e^{-[E_{\e i-j}(r-r_{\e i})+ E_{-\e i-j}(r-r_{-\e i})]/T }
	\nonum
           &  +&  .......
\enar
where the subscript \( i \) on \( Z \) indicates the partition function
in the presence of an infinitesimal \( i \) type vortex and \( V^{ii}_L \)
is the linearly screened potential between a vortex-antivortex pair of type
\( i \). The interaction energy
\( E_{\e i i}(r-r_{\e i}) \, = \, \d q_i q_i U_{ii}(r - r_{\e i}) \)
with similar definitions for the other interaction energies appearing in
eqs.(\ref{eq:4.9}) in terms of the interaction potentials \( U_{jj} \) and
\( U_{ij} \). According to our  ansatz for  the classical field
configuration \(\Phi\) of a nonabelian vortex configuration, is
\beq\label{eq:4.10}
       \Phi =
		{\cal G} \,\Phi_{diag}    \,{\cal G}^{-1}
\eeq
and the energy of such a configuration can be approximated by the kinetic
energy term in eq.(\ref{eq:2.1})
\beq\label{eq:4.11}
	E \approx
	        \int\, d^2 \br \,\tr \left[ g^{ab}\partial_a \Phi
		\partial_b \Phi   \right]
\eeq
Therefore, it is clear that the unscreened interaction energy
\(E^{0}_{\e ii}(\br_1-\br_{\e i}) \) between a background charge \(q_i\)
and infinitesimal charge \(\d\,q_i\) at points \(\br_i\) and \(\br_{\e i}\)
is given by
\bear\label{eq:4.12}
       	\Phi_{\e ii}(\br,\br_i,\br_{\e i})
		&=& {\cal G}_{\e i}\,{\cal
		G}_{i}\Phi_{diag} \,{\cal G}^{-1}_{ i}\,{\cal
		G}^{-1}_{\e\, i}
	\nonum
	E^{0}_{\e ii}(\br_i-\br_{\e i}
	&=& 	\int\, d^2
		\br \,\tr \left[ g^{ab}\partial_a \Phi_{\e ii}
		\partial_b \Phi_{\e ii} \right]
	\nonum
	&=&
		\int\, d^2
		\br \biggl\{ 2\e^2(\vf_1+2\,\vf_2)^2
		[\nabla_{r}A(\br-\br_{\e i})]^2
	\nonum
	&& \quad\quad +
		4\e(\vf_1+2\vf_2)^2 \nabla_{r} A(\br-\br_{\e i})
		\cdot\nabla_{r} A(\br-\br_i)
	\nonum
	&& \quad\quad
		+2(\vf_1+2\,\vf_2)^2 [\nabla_{r} A(\br-\br_i)]^2
		\biggr\}
	\nonum &=&
			2(\d q_{ i})^2 V_{0}^{i i}(0)+ 4\d
		q_{ i} \,q_i V_{0}^{ ii}(\br_i-\br_{\e i}) + 2 (
		q_i)^2 V_{0}^{ ii}(0)
\enar
where \( \d q_{ i} =\e\, q_i \) , \( A( \br ) \, = \, {\rm Im }\, [{\rm ln}
(r e^{i \theta} ) ]\) and
\bear\label{eq:4.13}
	V_{0}^{ii}(\br_i-\br_{\e i})
& =&
	\int  d^2 \br [\nabla_{r} A(\br -\br_{\e i})]\cdot
	[\nabla_{r}A(\br - \br _i)]
\nonum
\enar
is the unscreened interaction potential between two i-type
vortices. Furthermore,
\bear\label{eq:4.14}
	V_0^{i i }(0)
	&=& \int  d^2 \br [\nabla_{r} A (\br-\br_i)]^2  \cr
	 &  = &  \int  d^2 \br [\nabla_{r} A(\br - \br_{\e i})]^2
\enar
 are the self-energy of the background and test vortices at the points
\(\br\) and \(\br_{\e i}\) respectively. Since the test charge
\( \d q_{ i} \) is not thermally created, we should subtract it's
self-energy term \(2(\d q_{ i})^2 \, V_0^{\e ii}(0)\) from the right hand
side of eq.(\ref{eq:4.12}) Henceforth, we shall take for the form of the
interaction energy {\em including} the effects of screening, \( E_{ii} \),
the following
\beq\label{eq:4.15}
	E_{ ii}=  4\d q_{ i} q_i V_{L}^{ii}(\br_i-\br_{\e i})
		  + 2   q_i^2 V_{L}^{ ii}   (0)
\eeq
 ( This is analogous to the replacement of the unscreened logarithmic
 Coulomb potential by \( V_L \) in the previous section.)
 Similarly,
\bear\label{eq:4.16}
	E_{-\e ii}&=&  -4\d q_{ i} q_i V_{L}^{ii}(\br_i-\br_{-\e i})
		+ 2   q_i^2 V_{L}^{ ii}   (0) \nonum
	E_{ \e i-i}&=&  -4\d q_{ i} q_i V_{L}^{ ii}(\br_i-\br_{\e i})
      + 2   q_i^2 V_{L}^{ ii}   (0) \nonum
	E_{-\e i-i}&=&  4\d q_{ i} q_i V_{L}^{ii}(\br_i-\br_{-\e i})
		  + 2   q_i^2 V_{L}^{ ii}   (0)
\enar

Next we shall consider interaction energies between an \( i \) type test
vortex and a background vortex of type \(j\). This now involves 2 possible
orientations of the group elements \( {\cal G}_{\e \,i} \) and \( {\cal
G}_{j} \), since the latter elements are non-commuting, but a straight
forward calculation gives
\bear\label{eq:4.17}
\Phi_{\e ij}&=&
		   \,{\cal G}_{\e\, i}\,{\cal G}_{j}\Phi_{diag}
	           \,{\cal G}^{-1}_{ j}\,{\cal G}^{-1}_{\e\, i} , \qquad
	\Phi_{j \e i}=
		   \,{\cal G}_{j}\,{\cal G}_{\e \, i}\Phi_{diag}
	           \,{\cal G}^{-1}_{ \e \, i}\,{\cal G}^{-1}_{j}
	           \nonum
	E^{0}_{\e ij}&=&
		  {\displaystyle {1\over 2} }
		\{ \int\, d^2 \br \,\tr \left [g^{ab}\partial_a \Phi_{\e ij}
			\partial_b \Phi_{\e ij}
				   \right]
		+ \int\, d ^2 \br \,\tr \left [g^{ab}\partial_a \Phi_{j \e i}
			\partial_b \Phi_{j \e i}
				   \right] \}
	\nonum
	&=&
	   \int\, d^2 \br \biggl\{ \left( 2\,\e^2(\vf_1+2\,\vf_2)^2
	  - {3\over2}\e^2\vf_2(2\vf_1+\vf_2) \sin^2[A(\br-\br_j)]\right)
		    [\nabla_{r} A(\br-\br_{\e i})]^2
        \nonum
 	&+&
	   \left( 2(2\,\vf_1+\vf_2)^2  - {3\over2}\e^2\vf_2(2\vf_1+\vf_2)
		\sin^2[A(\br-\br_{\e i})] \right)
		    [\nabla_{r} A(\br-\br_{j})]^2
	         \biggr\}
	\nonum
\enar
whilst for a background anti-vortex of type \( j \) we have
 \bear\label{eq:4.18}
	\Phi_{\e i-j}&=&
		    \,{\cal G}_{\e\, i}\,{\cal G}^{-1}_{j}\Phi_{diag}
	            \,{\cal G}_{ j}\,{\cal G}^{-1}_{\e\, i}, \qquad
	\Phi_{-j \e i} =
		    \,{\cal G}^{-1}_{j}{\cal G}_{\e\, i}\Phi_{diag}
	            \,{\cal G}^{-1}_{\e\, i}{\cal G}_{ j}
\nonum
	E^{0}_{\e i-j}&=&
		\int\, d^2 \br \,\tr \left [g^{ab}\partial_a \Phi_{\e ij}
		\partial_b \Phi_{\e ij}  \right] +
		\int\, d^2 \br \,\tr \left [g^{ab}\partial_a \Phi_{-j \e i}
		\partial_b \Phi_{-j \e i}  \right]
\nonum
		&=&E^{0}_{\e i \, j }
\enar
{}From eqs.(\ref{eq:4.17}), (\ref{eq:4.18}) one can see that in fact, there
are no interaction terms between the two different type of charges \(\d q_{
i}\) and \(q_j\), which would have corresponded to terms {\em linear} in \(
\e \).  The terms present can be interpreted as the self energies of the \(
i \) type test charge and the
\( j \) type background charge. It should be noticed here that
the self energy of the test charge \(\d q_{ i}\) has been modified by the
\(\sin^2\) terms. Since the test charge is not thermally created, all such
self energy terms should be subtracted from \(E_{\e i j}\) as stated
previously.

As is shown in appendix A, the modification terms themselves do indeed take
the form of divergent self energies of the test charge \( \d q_i \) , so we
are justified in dropping them from the expressions for the interaction
energies.  Hence finally we have
\beq\label{eq:4.19}
	E_{ \e ij}=                    2   q_j^2 V_{L}^{j j} (0)
=E_{-\e ij}=  E_{ \e i-j}= E_{-\e i-j}
\eeq
 Similarly, one can check that
\bear\label{eq:4.20}
	E_{ \e jj}&=&   \ \ 4\d q_{ j} q_j V_{L}^{ jj}(\br-\br_{ \e j})
		+ 2   q_j^2 V_{L}^{ jj}   (0) \nonum
	E_{-\e jj}&=&      -4\d q_{ j} q_j V_{L}^{ jj}(\br-\br_{-\e j})
		+ 2   q_j^2 V_{L}^{ jj}   (0) \nonum
	E_{ \e j-j}&=&     -4\d q_{ j} q_j V_{L}^{ jj}(\br-\br_{\e j})
				+ 2   q_j^2 V_{L}^{jj}   (0) \nonum
	E_{-\e j-j}&=&  \ \ 4\d q_{ i} q_i V_{L}^{ jj}(\br-\br_{-\e j})
		  + 2   q_j^2 V_{L}^{jj}   (0)
\enar
 and
\beq\label{eq:4.21}
	E_{ \e ji}=  2   q_i^2 V_{L}^{ ii}  (0)  =
	E_{-\e ji}= E_{ \e j-i}= E_{-\e j-i}
\eeq
\noindent The density of positive(negative) background nonabelian
charges is defined as follows
\bear\label{eq:4.22}
	n_{ i}^{(\+-)}\,&=&\,\frac{e}{\pi\,\xi^2}^{(-4 V^{ ii}_L(0) -
		        E^{ i}_c)/T}\,
		\nonum
	         &=&\, \frac{z_{ i}}{\pi\,\xi^2}
				\,e^{-4 V^{ ii}_L(0)/ T}
		\nonum
	n_{ j}^{(\+-)}\,&=&\,\frac{e}{\pi\,\xi^2}^{(-  4 V^{jj}_L(0) -
		        E^{ j}_c)/T}
		\nonum
	         &=&\, \frac{z_{ j}}{\pi\,\xi^2}
				\,e^{-4 V^{jj}_L(0)/ T}
\enar
Substituting the expressions for the total energies \(E\) of the various
configuration as calculated above in terms of the linearly screened
potentials, we obtain for the partition function of eq.(\ref{eq:4.9})
\bear\label{eq:4.23}
 \lefteqn{ Z_{i} (\d  q_i, \br_{\e i},\br_{-\e i} ) \,=1+\int  d^2 \br  \
		e^{- \1/2\d q_{  i} \d q_{-  i} V^{ i i}_L(\br)
			  } \d^2 (\br -(\br_{\e i}-\br_{-\e i}))  }
              \nonum
	   \  & & \qquad +\, z_{i}^+
		     \int d^2 \br\  e^{
		     -[ 4\d q_{  i} q_{i}  V^{ ii}_L(\br-\br_{\e i})
	     	       -4\d q_{  i} q_{i}  V^{ ii}_L(\br-\br_{-\e i})
		       +4\, q_i^2 V^{ii}_L(0)]/T}
		\nonum
	    \  && \qquad+\, z_{i}^-
		     \int d^2 \br\  e^{
		     -[ -4\d q_{  i} q_{i}  V^{ ii}_L(\br-\br_{\e i})
	  	        +4\d q_{  i} q_{i}  V^{ ii}_L(\br-\br_{-\e i})
			+4\, q_i^2 V^{ii}_L(0)]/T}
		\nonum
	   \  & & \qquad+  \, z_{j}^+
		\int d^2 \br\  e^{ [-   4  q_{j}^2  V^{ jj}_L(0)]/T }
		\nonum
	   \  & & \qquad+  \, z_{j}^-
		\int d^2 \br\  e^{ [-   4  q_{j}^2  V^{ jj}_L(0)]/T }
              + ...
\enar
Rewriting the partition function in terms of the free charge
densities, we obtain
\bear\label{eq:4.24}
 Z_i \,&=& 1+\int  d^2 \br  \
		e^{ \1/2(\d q_{  i})^2   V^{ i i}_L(\br)
			  } \d^2 (\br -(\br_{\e i}-\br_{-\e i}))
              \nonum
	   && \quad +  \, n_{i}^+
			\int d^2 \br\  e^{ -  4\d q_{  i} q_{i}
			[ V^{ ii}_L(\br-\br_{\e i})
			 -  V^{ ii}_L(\br-\br_{-\e i})]/T}
		\nonum
	   && \quad +  \, n_{i}^-
			\int d^2 \br\  e^{ -4\d q_{  i} q_{i}
			[-V^{ ii}_L(\br-\br_{\e i})
		 +   V^{ii}_L(\br-\br_{-\e i})]/T}      + ...
\enar
\sect{ P-B equations for nonabelian vortices}

Before we compute the Poisson-Boltzmann equations of the system, we shall
consider the relationship between the two different types of nonabelian
vortex charge. The energy density (again dropping the potential term as
explained earlier) of a single \(i\) type vortex is written as
\beq\label{eq:5.1}
	{\cal E}_i = \tr \left [ g^{ab} \ld{a} \Phi_i \ld{b}\Phi_i \right ]
\eeq
which, up to core corrections, can be approximated  by
\beq\label{eq:5.2}
	{\cal E}_i =
	\frac{1}{2\,r^2} ( (\vf_1+2\vf_2) \nabla_{r} A[\br - \br_1]{)}^2
\eeq
at distance scales greater than the typical vortex core size \( \xi \).
In eq.(\ref{eq:5.2}) we have considered a single vortex of type \(i\)
located at the point
\(\br_1\)  and the field configuration \(\Phi_i\) is given by
\[ \Phi_i= \cG_i\Phi_\diag \cG_{-i}\] as described earlier.

{\def\strut{\vrule height 6ex depth 6ex width 0pt}
\footnotesize
\begin{table}
 \begin{center}
 \begin{tabular}{|l|l|c|c|}
 \hline
 \(\strut\qquad\qquad U \quad\)&\(\tilde{{\cal G}_i}=U \cG_i U^{-1} \) &
 \(\tilde{{\cal G}_j}=U \cG_j U^{-1} \)&\( U<\Phi_{diag}>U^{-1} \)
 \\ \hline
 \(\strut
 \pmatrix{-1 &  0 & 0 \cr
	   0 & -1 & 0 \cr
	   0 &  0 & 1 \cr  } \)
				&
				\(\qquad  {\cal G}_{-i} \)
						&
						\(  {\cal G}_{-j} \)
 & \( \pmatrix{\vf_1 &  0     & 0 \cr
		0    & \vf_2 & 0 \cr
		0    &  0     & -(\vf_1+\vf_2) \cr  } \)
						\\
 \(\strut
 \pmatrix{-1 &  0 & 0 \cr
	   0 &  1 & 0 \cr
	   0 &  0 &-1 \cr  } \)
				&
				\(\qquad   {\cal G}_{-i} \)
						&
						\(  {\cal G}_{j} \)
 & \( \pmatrix{\vf_1 &  0     & 0 \cr
		0    & \vf_2 & 0 \cr
		0    &  0     & -(\vf_1+\vf_2) \cr  } \)
						\\
 \(\strut
 \pmatrix{ 0 &  1 & 0 \cr
	   1 &  0 & 0 \cr
	   0 &  0 & 1 \cr  } \)
				&
				\(\qquad {\cal G}_{j} \)
						&
						\(  {\cal G}_{i} \)
 & \( \pmatrix{\vf_2 &  0     & 0 \cr
		0    & \vf_1 & 0 \cr
		0    &  0     & -(\vf_1+\vf_2) \cr  } \)
						\\
 \hline
 \end{tabular}
 \end{center}
 \caption{}
 \label{tab2}
 \end{table}
}
Similarly, for a single  \(j\)-type vortex at the point  \( \br_2 \),
\beq\label{eq:5.3}
	{\cal E}_j = \tr \left [ g^{ab} \ld{a} \Phi_j \ld{b}\Phi_j \right ]
\eeq
\beq\label{eq:5.4}
	{\cal E}_j =
	\frac{1}{2\,r^2} ( (2\vf_1+\vf_2) \nabla_{r}  A[\br -\br_2] {)}2
\eeq
{}From table 1, it is clear that we can transform \(\cG_i\) to \( \cG_j\) by
using the constant transformation matrix \(U\), so we can rewrite
\(\Phi_j\) in terms of \(\cG_i\) as follows
\beq\label{eq:5.5}
 \Phi_j= U\cG_iU^{-1}\Phi_\diag U^{-1} \cG_{-i}U
\eeq
Thus we could have obtained the energy density of the \(j\)-type vortex, by
substituting using eq.(\ref{eq:5.5} ) into the self-energy of an \( i \)
type vortex.  Since we know that \(U\) is unitary, by the property of the
trace which is invariant under the similarity transformation
\[ \tr[B]=\tr[UBU^{-1}]\]
we then conclude that
\beq\label{eq:5.6}
  \tr \left [ g^{ab} \ld{a} \Phi_j \ld{b}\Phi_j \right ] = \tr \left [
g^{ab} \ld{a} \Phi_i \ld{b}\Phi_i \right ]_{\vf_1\leftrightarrow\vf_2}
\eeq
which is evident from comparing \( {\cE}_i \) with \( { \cE}_j \) .  It
 follows that the infinitesimal charges \(\d q_i\) and \(\d q_j\) transform
 into one another by simply interchanging \(\vf_1\) and \(\vf_2\) in
 \(\Phi_\diag\).  This is sufficient to show that both infinitesimal
 charges \(\d q_i\) and \(\d q_j\) depend on the {\em same} infinitesimal
 rotation parameter \(\e\).

To find the linear screening potential of the system with both types of
infinitesimal charge \(\d q_i\) and \(\d q_j\) present, we simply use the
chain rule by differentiating the partition function first with respect to
\(\e\) and apply the following relations
\bear\label{eq:5.7}
	\d q_i&=& \frac{\e}{\sqrt{2}} (\vf_1+2\,\vf_2)
	\nonum
	\frac{\pd \d q_i}{\d \e} &=&
		\frac{1}{\sqrt{2}}(\vf_1+2\,\vf_2)\,=\, q_i
	\nonum
	\d q_j&=& \frac{\e}{\sqrt{2}} (2\,\vf_1+\vf_2)
	\nonum
	\frac{\pd \d q_j}{\d \e} &=&
		\frac{1}{\sqrt{2}}(2\,\vf_1+\vf_2)\,=\, q_j
\enar
For later use, we shall need to consider the case when infinitesimal test
charges of type \(i\) and \(j\) are simultaneously present in the
system. For this system with two different test charges \(\d q_{i}\) and
\(\d q_{ j}\) at \(\br_{\e i}\) and \(\br_{\e j}\), the lowest order
partition function can be reduced to
\bear\label{eq:5.8}
 Z_{ij} ( \d q_i  , \d q_j  ,  \br_{\e i}  , \br_{\e j} )\,
	   &=&
		1+\int  d^2 \br
		 \ e^{-\1/2\d q_{ i} \d q_{ j} V^{i j}_L(\br)/T}
			   \d^2 (\br-(\br_{\e i}-\br_{\e j}))
              \nonum
	   && \quad +z_i              \int d^2 \br\
			e^{-[E_{\e ii}+E_{\e ji} ]/T }
              \nonum
	   && \quad +z_i              \int d^2 \br \
			e^{-[E_{\e i-i}+ E_{\e j-i} ]/T }
	      \nonum
	   && \quad +z_j              \int d^2 \br\
			e^{-[E_{\e ij}+E_{\e jj} ]/T }
              \nonum
	   && \quad +z_j              \int d^2 \br \
			e^{-[E_{\e i-j}+ E_{\e j-j} ]/T }
             + ...
\enar
where the subscript \( ij \) on \( Z\) in eq.(\ref{eq:5.8}) denotes the
fact that both \( i \) and \( j \) type test charges are present.
Substituting the expressions for the total energy \(E\) from the previous
section, we get
\bear\label{eq:5.9}
  Z_{ij} ( \d q_i  , \d q_j  ,  \br_{\e i}  , \br_{\e j} ) \,
	   &=& 1+\int  d^2 \br  \
		e^{- \1/2 \d q_{ i} \d q_{ j} V^{ i j}_{L} (\br)/T
			  } \d^2 (\br -(\br_{\e i}-\br_{\e j}))
              \nonum
	   && \quad +  \, n_{i}^+
			\int d^2 \br\  e^{
			-  4\d q_{ i} q_{i}  V^{ ii}_{L} (\br-\br_{\e i}) /T}
		\nonum
	   && \quad +  \, n_{i}^-
			\int d^2 \br\  e^{
			 4\d q_{ i} q_{ i}  V^{ ii}_{L}(\br-\br_{\e i}) /T}
		\nonum
	   && \quad +  \, n_{j}^+
			\int d^2 \br\  e^{
			 4\d q_{ j} q_{j}  V^{ jj}_{L}(\br-\br_{\e j}) /T}
		\nonum
	   && \quad +  \, n_{j}^-
			\int d^2 \br\  e^{
			 -4\d q_{ j} q_{j}  V^{ jj}_{L}(\br-\br_{\e j}) /T}
	     + ...
\enar
The linearly screened potential at (\(\br_{\e i}-\br_{\e j}\)) is given by
\beq\label {eq:5.10}
	\d q_{ i} V_{L}^{ ij}(\br_{\e i}-\br_{\e j})
	  =T \frac{\partial \, \ln\,Z_{ij}}{\partial \,(\d q_{ j})}
	  =\,{T\ovr Z_{ij}}
		\frac{\partial Z_{ij}}{\partial \,(\d q_{ j})}
\eeq
and
\beq\label{eq:5.11}
	\d q_{ j} V^{ ji}_L(\br_{\e j}-\br_{\e i})
	  =T \frac{\partial \, \ln\,Z_{ij}}{\partial \,(\d q_{ i})}
	  =\,{T\ovr Z_{ij}}
		\frac{\partial Z_{ij}}{\partial \,(\d q_{ i})}
\eeq
Now we are in a position to construct a Poisson-Boltzmann equation for the
linearly screened potentials between various test charges. In the first
case, two test charge will be taken to be of the same type, whilst in the
second case we will take different types.

For the same type of test charges, say \(\d q_{ i}\) and \(\d q_{- i}\),
the linearly screened potential at (\(\br_{\e i}-\br_{-\e i}\)) is given
by
\beq\label {eq:5.12}
	\d q_{  i} V_{L}^{ ii}(\br_{\e i}-\br_{-\e i})
	  =T \,\frac{\partial \, \ln\,Z_{i}}{\partial \,(\d q_{  i})}
	  =\,{T\ovr Z_{i}}
		\frac{\partial Z_{i}}{\partial \,(\d q_{  i})}
\eeq
Since we use the two infinitesimal charges \(\d q_{ i},\,\d q_{j}\), we
can ignore the higher order terms and therefore approximate \(1/Z_{i} \) by
1 (\( Z_{i}\,=\,1+O(z_i, z_j )+O(\d q_{ i}^2)+O(\d q_{ i}^2)\)). Hence we
find
\bear\label{eq:5.13}
\lefteqn{\d q_{  i} V_L^{ ii}(\br_{\e i}-\br_{-\e i}) \ = \
		 \,\d q_{  i} \,V_L^{ i i}(\br_{\e i}-\br_{-\e i})\,
		 e ^{\1/2(\d q_{  i})^2\,\,
			V_L^{ i i}(\br_{\e i}-\br_{-\e i})/T} }
		\nonum
	    & - {\displaystyle
		      \,\frac{4 q_i\,n_{i}^+}{ \,T}
			\int }d^2 \br \ [ V_L^{ ii}(\br-\br_{\e i})-
			V_L^{ ii}(\br-\br_{-\e i})]
			e^{ -  4\d q_{  i} q_{i}
			[ V^{ ii}_L(\br-\br_{\e i})
			 -  V^{ ii}_L(\br-\br_{-\e i})]/T}
             \nonum
	    & - {\displaystyle
		       \,\frac{4 q_i\,n_{i}^-}{ \,T}
			\int} d^2 \br \ [-V_L^{ ii}(\br-\br_{\e i})
			+ V_L^{ ii}(\br-\br_{-\e i})]
			e^{ -  4\d q_{  i} q_{i}
			[ -V^{ ii}_L(\br-\br_{\e i})
			 + V^{ ii}_L(\br-\br_{-\e i})]/T}
	    \nonum
\enar
To lowest order \(V_L^{ ii}(\br) = V_0^{ ii}(\br) + O(z_i , z_j )\) where
the unscreened potential \( V_0^{ ii}(\br) \) varies logarithmically with
distance.  Applying the Laplacian operator \(\nabla^2_{ {\bf r}_0} \) to
both sides of eq.(\ref{eq:5.13}) where \({ {\bf r}_0} \) is the relative
position of test charges \({ {\bf r}_0}=\br_{\e i}-\br_{-\e i}\), we get
\bear\label{eq:5.14}
\lefteqn{\d q_{  i}\nabla^2_{ {\bf r}_0} V_L^{ ii}(\br_{\e i}-\br_{-\e i})
		  \    = \
		 \,\d q_{  i} \nabla^2_{ {\bf r}_0}
			\,V_L^{ii}(\br_{\e i}-\br_{-\e i})\,
		 e ^{\1/2(\d q_{  i})^2\,\,
			V_L^{ ii}(\br_{\e i}-\br_{-\e i})} }
		\nonum
	    & -       {\displaystyle
			 \,\frac{4 q_i\,n_{i}^+}{ \,T}
			\int }d^2 \br \nabla^2_{ {\bf r}_0}
			\ [ V_L^{ ii}(\br-\br_{\e i})-
			V_L^{ ii}(\br-\br_{-\e i})]
			e^{ -  4\d q_{  i} q_{i}
			[ V^{ ii}_L(\br-\br_{\e i})
			 -  V^{ ii}_L(\br-\br_{-\e i})]/T}
             \nonum
	    & -   {\displaystyle
			\,\frac{4 q_i n_{i}^-}{ \,T}
			\int }d^2 \br \nabla^2_{ {\bf r}_0}
			 [-V_L^{ ii}(\br-\br_{\e i})+
			V_L^{ ii}(\br-\br_{-\e i})]
			e^{ -  4\d q_{  i} q_{i}
			[ -V^{ ii}_L(\br-\br_{\e i})
			 + V^{ ii}_L(\br-\br_{-\e i})]/T}
	\nonum
\enar
Since, as mentioned, the unscreened potential
\(V^{ii}_0(\br_{\e i}-\br_{-\e i})\)  is proportional
to the logarithm function, it follows that
\beq\label{eq:5.15}
\nabla^2_{\br_{0}}\,V^{\e ii}_0(\br-\br_{\e i})
			  =\,-2\pi \,\d^2 (\br-\br_{\e i})
\eeq
Hence, using eqs.(\ref{eq:5.14}), (\ref{eq:5.15}) we find a
Poisson-Boltzmann equation satisfied by \mbox{\( V^{ ii}_L(\br_{\e
i}-\br_{-\e i})
\)}
\bear\label{eq:5.16}
 \d q_{  i}\nabla^2_{ {\bf r}_0} V_L^{ ii}(\br_{\e i}-\br_{-\e i})
		  &  = &
		 -2\pi  \,\d q_{  i} \d^2(\br_{\e i}-\br_{-\e i})\,
		\nonum
	    & & +      \,{8\pi  q_i\,n_{i}^+ }
			\,e^{  4\d q_{  i} q_{i}
			[ V^{ ii}_L(\br_{\e i}-\br_{-\e i})
			 -  V^{ ii}_L(0)]/T}]
             \nonum
	    & & -      \,{8\pi q_i\,n_{i}^-}
			\, e^{ -  4\d q_{  i} q_{i}
			[  V^{ ii}_L(\br_{\e i}-\br_{-\e i})
			 - V^{ ii}_L(0)]/T}
\enar
Again as in the abelian case we redefine
\beq \label{eq:5.17}
	V^{ ii}_L(\br_{\e  i}-\br_{-\e i}) - V^{  ii}_L(0)
\to {\tilde V}^{ ii}_L(\br_{\e i}-\br_{-\e i})
\eeq
We get finally,  (dropping \~ \ (tilde)\ )
\beq\label{eq:5.18}
	 \nabla^2_{\br_0} V^{ ii}_L(\br_{\e i}-\br_{-\e i})
	\,=\,-   2\pi \,     \,\d^2(\br_{\e i}-\br_{-\e i})
             -\,\frac{32 \pi \,\,{\a}_{i} \,n_i}{T}
			 V^{ ii}_L(\br_{\e i}-\br_{-\e i})
\eeq
where \( n_{i}^++n_{i}^-\,=\, n_{i}\) and \( {\a}_{i} \, = \, {q_{i}}^2 \)
is the analogue of the quantity \( \a \) introduced in section 2.

Similarly, for the two test charges of \(j\) and \(-j\) type, we can derive
a P-B equation for the corresponding linearly screened potential
\( V^{ jj}_L(\br_{\e j}-\br_{-\e j})\)
\beq\label{eq:5.19}
	 \nabla^2_{\br_0} V^{ jj}_L(\br_{\e j}-\br_{-\e j})
	\,=\,-   2\pi   \,\d^2(\br_{\e j}-\br_{-\e j})
             -\,\frac{32 \pi \,\a_{j}\,n_j}{T}
			 V^{ jj}_L(\br_{\e j}-\br_{-\e j})
\eeq
with \( n_{j}^++n_{j}^-\,=\, n_{j}\) and \( {\a}_{j} \, = \, {q_{j}}^2 \)
It should be remembered that in deriving the P-B eqs.(\ref{eq:5.16}) and
(\ref{eq:5.19} ), we have used the property, that
\( V_L^{i j} \) is vanishing at lowest order, as discussed in section 4
immediately following eqs.(\ref{eq:4.18}).  This is why \( V_L^{ i j} \)
plays no role in the P-B equations derived above.  A consistency condition
on this property of the screened potential can be derived by considering a
P-B equation for \( V_L^{ i j} \) itself, and showing that the solution to
lowest order is indeed vanishing.  To derive such a P-B equation, we
consider a simultaneous configuration of different types of infinitesimal
test charges \(\d q_{ i},\,\d q_{ j}\). From (\ref{eq:5.10}) we obtain
\bear\label{eq:5.20}
\d q_{ i} V^{ i j}_{L}(\br_{\e i}-\br_{\e j})
	 &=& \frac{T}{q_i}\frac{\pd \ln Z_{ij}}{\pd \e}
	\nonum
   & = &		 -\,\d q_{ i} \,V_L^{i j}(\br_{\e i}-\br_{\e j})\,
		 e ^{-\1/2 \d q_{ i}\d q_{ j} \,
		\,V_L^{ i j}(\br_{\e i}-\br_{\e j})/T}
		\nonum
	    && -       \,\frac{4 q_i^2\,n_{i}^+}{ q_j}
			\int d^2 \br \   V_L^{ ii}(\br-\br_{\e i})
			e^{ -  4\d q_{ i} q_{i}
			  V^{ ii}_L(\br-\br_{\e i}) /T}
		\nonum
	    && +       \,\frac{4 q_i^2\,n_{i}^-}{ q_j}
			\int d^2 \br \   V_L^{ ii}(\br-\br_{\e i})
			e^{    4\d q_{ i} q_{i}
			  V^{ii}_L(\br-\br_{\e i}) /T}
		\nonum
	    && +       \,\frac{4 q_j^2\,n_{j}^+}{ q_j}
			\int d^2 \br \   V_L^{ jj}(\br-\br_{\e j})
			e^{ -  4\d q_{ j} q_{j}
			  V^{jj}_L(\br-\br_{\e j}) /T}
		\nonum
	    && -       \,\frac{4 q_j^2\,n_{j}^-}{ q_j}
			\int d^2 \br \   V_L^{ jj}(\br-\br_{\e j})
			e^{    4\d q_{ j} q_{j}
			  V^{jj}_L(\br-\br_{\e j}) /T}
		\nonum
	    && + \quad ...
\enar
To lowest order in \( z_i \) and \( z_j \),
\beq\label{eq:5.21}
	V_L^{ij}(\br) = V_0^{ij}(\br) + O( z_i , z_j )
\eeq
where as we have shown in Appendix A, \( V_0^{ ij}(\br) \) is in fact a
divergent self energy contribution to the test charge, independent of \(
\br \).  Applying the Laplacian operator \(\nabla^2_{ {\bf r}_0} \) to both
sides of eq.(\ref{eq:5.20}), where \({ {\bf r}_0} \) is the relative
position of the test charges \({ {\bf r}_0}=\br_{\e i}-\br_{\e j}\), we
obtain
\bear\label{eq:5.22}
 \d q_{ i} \nabla^2_{\br_{0}}V^{i j}_L(\br_{\e i}-\br_{\e j})
	    & = &
		 -\,\d q_{ i} \,\nabla^2_{ {\bf r}_0}
			V_L^{ i j}(\br_{\e i}-\br_{\e j})\,
		 e ^{-\1/2 \d q_{ i}\d q_{ j} \,\,V_L^{ i j}
			(\br_{\e i}-\br_{\e j})/T}
		\nonum
	    && -       \,\frac{4 q_i^2\,n_{i}^+}{ q_j\,T}
			\int d^2 \br \
			\left[ \nabla^2_{ {\bf r}_0}
				V_L^{ii}(\br-\br_{\e i})
			\right]
			e^{ -  4\d q_{ i} q_{i}
			  V^{ii}_L(\br-\br_{\e i}) /T}
		\nonum
	    && +       \,\frac{4 q_i^2\,n_{i}^-}{ q_j\,T}
			\int d^2 \br \
			\left[ \nabla^2_{ {\bf r}_0}
					V_L^{ii}(\br-\br_{\e i})
			\right]
			e^{    4\d q_{ i} q_{i}
			  V^{ ii}_L(\br-\br_{\e i}) /T}
		\nonum
	    && +       \,\frac{4 q_j^2\,n_{j}^+}{ q_j\,T}
			\int d^2 \br \
			\left[ \nabla^2_{ {\bf r}_0}
				V_L^{ jj}(\br-\br_{\e j})
			\right]
			e^{    4\d q_{ j} q_{j}
			  V^{ jj}_L(\br-\br_{\e j}) /T}
		\nonum
	    && -       \,\frac{4 q_j^2\,n_{j}^-}{ q_j\,T}
			\int d^2 \br \
			\left[ \nabla^2_{ {\bf r}_0}
				V_L^{ jj}(\br-\br_{\e j})
			\right]
			e^{   - 4\d q_{ j} q_{j}
			  V^{ jj}_L(\br-\br_{\e j}) /T}
		\nonum
	     && + ...
\enar
{}From the properties of the lowest order potentials \( V_0^{ ii} \, ,
V_0^{ jj} \) and \( V_0^{ ij} \) discussed above,
the P-B equation becomes
\bear\label{eq:5.23}
 \d q_{ i} \nabla^2_{\br_{0}}V^{ i j}_L(\br_{\e i}-\br_{\e j})  & = &
	   \frac{8\pi q_i^2\,n_{i}^+}{ q_j}
			e^{ -  4\d q_{ i} q_{i}
			  V^{ ii}_L(0) /T}
		\nonum
	    && -       \,\frac{8\pi q_i^2\,n_{i}^-}{ q_j}
			e^{    4\d q_{ i} q_{i}
			  V^{ ii}_L(0) /T}
		\nonum
	    && -       \,\frac{8\pi q_j^2\,n_{j}^+}{ q_j}
			e^{    4\d q_{ j} q_{j}
			  V^{jj}_L(0) /T}
		\nonum
	    &&+        \,\frac{8\pi q_j^2\,n_{j}^-}{ q_j}
			e^{ - 4\d q_{ j} q_{j}
			  V^{ jj}_L(0) /T}
\enar
For neutral system, we know that \(n^+_i=n^-_i\), then expanding
the exponential terms out to the lowest order in fugacities, we get the final
form of the P-B equation for \( V_L^{ ij} \)
\bear\label{eq:5.24}
 \nabla^2_{ {\bf r}_0} V_L^{ ij}(\br_{\e i}-\br_{\e j})
		  &  = &
	    -      \frac{32\pi \a_{i} q_i \,n_{i} }{ q_j\,T}
			\,    V^{ii}_L(0)
             \nonum
	    & & -      \,\frac{32\pi\a_{j} \,n_{j}} { T}
			   V^{jj}_L(0)
\enar
A similar equation can be derived for \( V_L^{ji} \) which is given by the
r.h.s. of eq.(\ref{eq:5.24}) but with the subscripts \( i \) and \( j \)
interchanged.  Notice that an important difference between
eq.(\ref{eq:5.24}) and the previous two P-B equations, (\ref{eq:5.18}) and
(\ref{eq:5.19}) is the absence of the (fugacity independent) delta function
term on the r.h.s.. Indeed the r.h.s. of eq.(\ref{eq:5.24}) is simply
constant in the variable \( {\bf r}_0 \)\footnote{Thus the vanishing of
\(V_L^{ij} \)found previously to lowest order in fugacity is consistent with
the P-B equation for \(V_L^{ij}\)}.  Before moving on to discuss the
solutions and consequences of these equations, it is clear that the role of
the potential \( V_L^{ij} \) can be ignored, to the order in which we work.

A comparison of eqs.(\ref{eq:5.18}) and (\ref{eq:5.19}) to that of the P-B
equation derived in section 2 for the abelian Coulomb gas, shows that the
potentials \( V_L^{i i} \) and \( V_L^{j j} \) are essentially Coulombic in
nature with the free charge density and fugacity \( n_F \) and \( z \)
replaced by the same quantities with subscripts \( i , j \) appropriate for
the system of nonabelian vortices. Consequently, the solutions of
eqs.(\ref{eq:5.18}) and (\ref{eq:5.19}) are exactly of the type described in
section 3, with the replacement of fugacity \( z \) by \( z_i \, , z_j \)
and screening length \( \omega \) by \( \omega_{i}(z_i , T) \) and \(
\omega_{j}(z_j , T) \).  Furthermore we see that there will in principle
be two lines of critical temperatures \( T_{c}^{i} \) and \( T_{c}^{j} \)
given by
\bear\label{eq:5.25}
   T_{c}^{i} \,&  =& {{\a_{i}}\ovr 4}( 1 - 2 \pi z_i )  \,
   \nonum
    T_{c}^{j}\, &= & {{\a_{j}}\ovr 4} ( 1 - 2 \pi z_j )
\enar
It is perhaps worth commenting at this point, that the two critical
temperatures given in eqs.(\ref{eq:5.25} ) for vanishing fugacities
\(z_i \) and \( z_j \) , are in agreement with those one could have
obtained through heuristic free energy arguments  first discussed in
[6].
If we imagine increasing the temperature of the system from \( T \, = \, 0
\), eqs.(\ref{eq:5.25}) indicate that either the \( i \) followed by the
\( j \) type vortices dissociate or vice versa depending on the relative
magnitude of the critical temperatures in eq.(\ref{eq:5.25}). Note that
since the quantities \( \a_i \) and \( \a_j \) can be re-expressed in terms
of the vacuum values of the fields \( \vf_1 \) and \( \vf_2 \)
\bear\label{eq:5.26}
    \a_{i} \,& =&  \,\1/2 {< \vf_1 +2 \vf_2 > }^2 \cr
    && \cr
    \a_{j} \,& =&  \,\1/2 {< \vf_2 +2 \vf_1 > }^2
\enar
There is a constraint on \( \a_{i}\) and \( \a_{j} \) since the vacuum
values of \( \vf_1 \, , \vf_2 \) must lie on the ellipse
\( \vf_1 ^2 + \vf_2^2 + \vf_1 \vf_2 \,
	= \, {{\mu_{0} }^2} /{( \lambda + \lambda^{'} )}\) .
This constraint in terms of \( \a_{i}\) and \( \a_{j} \) is
\beq\label{eq:5.27}
	\a_{i} + \a_{j} - \sqrt{\a_{i} \a_{j} } \, = \,
	{ {3{\mu_{0}^2 } }\ovr{2 (\lambda + \lambda^{'} ) }  }
\eeq
which shows that the values of \( \sqrt{\a_{i}}, \sqrt{\a_{j}} \) also
parameterize an ellipse. It is clear from this that there exists an ellipse
of critical temperatures \( T_c^i \) and \( T_c^j \),  as the
fugacities \( z_i \) and \( z_j \) are taken to zero,
\beq\label{eq:5.28}
 T_{c}^{i} + T_{c}^{j} - \sqrt{T_{c}^{i} T_{c}^{j} } \, = \,
	{ {3{\mu_{0}^2 } }\ovr{2 (\lambda + \lambda^{'} ) }  }
\eeq
 This curious phase
structure is clearly a consequence of the degeneracy present in the vacuum
manifold which as we have seen in section 2, is an essential feature of the
nonabelian vortices considered in this paper.

In conclusion, we have investigated in this paper lowest order
screening in a system of nonabelian vortices in 2-dimensions,
through a detailed study of Poisson-Boltzmann like equations
for the various inter-vortex potentials.
Remarkably, we have seen that such equations are similar in
form to those of an abelian system of vortices , the latter being
physically equivalent to a Coulomb gas. This feature might be
a consequence of the simple model  of nonabelian vortices
we studied and perhaps also
to the fact that  we considered only the lowest non-trivial order
in a small fugacity expansion. It would certainly be interesting
to verify if these features persist in other models of nonabelian
vortices, involving different homotopy groups. In the present model,
we find a Kosterlitz-Thouless type phase structure in the underlying
system, i.e. both \( i \) and \( j \) type vortices undergo K-T like
phase transitions at some  particular critical temperatures.
What is interesting, and
a direct consequence of the nonabelian nature of the system, is
the connection between these critical temperatures, as evident
in eq.(\ref{eq:5.28}).
\newpage
{\Large{\bf Acknowledgements}}
\vskip0.1in
The work of  S.T.  was supported by the Royal Society
of Great Britain.
\vskip1in
\appendix
\begin{flushleft}
{\Large {\bf Appendix A }}
\end{flushleft}
\section*{Self Energy Modifications}
\setcounter{equation}{0}
\renewcommand{\theequation}{A.\arabic{equation}}

In this section we wish to show that the term
\(V_M\).
\beq\label{eq:A.1}
	V_M=\int d^2 z \sin^2[A(z-z_2)] \ld{z}A(z-z_1)\ld{\zb}A(z-z_1)
\eeq
(see eqs.(4.17) and (4.18) ), can indeed be interpreted as
modifications to the divergent self energy of test charges placed in the
system, as claimed in section 4. In eqn. (A.1) \( z_1,\,z_2 \) are the
positions of the \(i\) type test charge and \( j \) type
background charge  respectively.
Rather than compute \(V_M\) by integrating,  we instead consider  \(\nabla^2
\,V_M\) since this quantity appears in the corresponding  P-B  equation. One
finds
\bear\label{eq:A.2}
\ld{z_1}\ld{\zb_1} V_M
  &=& \ld{z_1} \int d^2 z \sin^2[A(z-z_2)] \nonum
  &&  \qquad \times
	\biggl[ \ld{\zb_1}\ld{z}A(z-z_1)\ld{\zb}A(z-z_1)
	+ \ld{z}A(z-z_1)\ld{\zb_1}\ld{\zb}A(z-z_1)
	\biggl]
  \nonum
  &=& \ld{z_1} \int d^2 z \sin^2[A(z-z_2)]
	\big[ \ld{z_1}\d^2 (z-z_1)\frac{1}{\zb-\zb_1}
	      +\ld{\zb_1}\d^2 (z-z_1)\frac{1}{z-z_1}
  \nonum && \hskip 4 cm
	      + \d^2 (z-z_1) \d^2 (z-z_1)
	      + \frac{1}{|z-z_1|^4}
	\big]
\enar
where  we remind the reader that  \( A(z-z_1 ) \, = \, {\rm Im}({\rm ln} (z-z_1
)) \)

 One can see  that all terms  in eq.(A.2) are manifestly singular except the
last term, and hence may be subtracted from the Hamiltonian of the
system as discussed in section 4. Let us concentrate then on the remaining
term,
 \beq\label{eq:A.3}
	\int d^2 z \frac{\sin^2[A(z-z_2)]}{|z-z_1|^4}
 \eeq
 By  a change of  variables
 \beq\label{eq:A.4}
	\sin^2[A(z'+\o_i)]
	   \ =\  {1\ovr 2}\,-\,{1\ovr 4}\,
	   \Biggl[
	   \biggl( \frac{z'+\o_i}{\zb'+\ob_i} \biggl)
	  +\,\biggl( \frac{\zb'+\ob_i}{z'+\o_i}  \biggl)
	    \Biggl]
\eeq
we obtain
\bear\label{eq:A.5}
\int \frac{d^2 z}{|z-z_1|^4}
		\left\{ \1/2 \,-\,\frac{1}{4}\,
			   \left[
				\biggl( \frac{z-z_2}{\zb-\zb_2} \biggl)
			     +\,\biggl( \frac{\zb-\zb_2}{z-z_2}  \biggl)
			   \right]
		\right\}
	\nonum \ =\
	\int \frac{d^2 z}{|z-z_1|^4}
		\left\{ {1\ovr 2}\,-\,{1\ovr 4}\,
		   \left[
		      \frac{2\Re[zz]+2\Re[z_2 z_2]-4\Re[z z_2] }
		           {|z-z_2|^4}
		   \right]
		\right\}
\enar
 Next, we wish to consider the behaviour of the
integral as \(  z \rightarrow z_1 \)  . In this respect we define
\bear\label{eq:A.6}
       z'    &=&   \,z-z_1  \ =\  r'e^{i\theta} \nonum
              \D    &=&  z_1-z_2  \ =\  d \cos\b + i\, d \sin\b \nonum
\enar
so that  we get for the integral in eq.(A.3)
\beq\label{eq:A.7}
  \int  \frac{d^2 r'}{r'^4}
	\left\{ {1\ovr 2}\,-\,{1\ovr 4}\,
	   \left[
      \frac{2\,r'^2\cos 2\theta +2\,d^2\cos 2\b-4\, r'd\cos(\theta+\b) }
	           {(r'+d)^4}
		   \right]
		\right\}
\eeq
 which has  divergent contributions when \(z\to z_1 \). In fact the
\( 1/ {r'}^4 \) term in this expression is exactly what we would
expect to obtain if we act on the (usual) logarithmically divergent
self energy of a single vortex in 2-dimensions, with \(\nabla^2 \) .

Having shown that \(\ld{z_1}\ld{\zb_1} V_M\) gives rise to divergent terms
we can conclude that contributions like \( V_M \) can be interpreted as
self energy modifications to the test charges, and hence should be
subtracted from the effective hamiltonian of our system.
\vspace{2ex}
\newpage
\begin{center}
{\large References }
\end{center}
\vspace{2ex}
\begin{description}
\item{[1]} P. Minnhagen, {\it Rev. Mod. Phys.} {\bf 59} (1987) 1001
\item{[2]} A. Vilenkin , {\it Phys. Rep.} {\bf 121} (1985) 263
\item{[3]} L. Onsager, {\it Nuovo Cimento } {\bf 6 }, {\it Suppl.} {\bf 2}
		(1949) 249; R.P. Feynmann, {\it Progress in low temperature
		physics}, {\bf Vol.1} ed. C.J. Gorter (North-Holland,
		Amsterdam, 1955) p.17; A.L. Fetter,
		{\it Phys. Rev. Lett.} {\bf 27} (1971) 986,
		{\it Ann. Phys.} (N.Y.) {\bf 70} (1972) 67
\item{[4]} L.P. Kadanoff, {\it J. Phys} {\bf A11} (1978), 1399
\item{[5]} V.L. Berezinsky, {\it Zh. Eskp.
	Teor. Fiz. } {\bf 61} (1972) 1144  [{\it Sov. Phys. JETP } {\bf 34 }(1972)
	610 ]
\item{[6]} J.M. Kosterlitz and D.J. Thouless, {\it J. Phys.
	}{\bf C6 }(1973)1181
\item{[7]} D.R. Nelson and J.M. Kosterlitz,
	{\it Phys. Rev. Lett. }{\bf 39 }(1977) 1201
\item{[8]} D.C. McQueeny, G. Agnolet and J.D. Preppy,
	{\it  Phys. Rev.Lett. }{\bf 52 }(1984) 1325
\item{[9]} C. Kobdaj and S. Thomas, {\it Nucl. Phys.}
         {\bf B413}  (1994) [FS] 689
\item{[10]} J. Fr\"{o}lich, in {\it Renormalization
	Theory}, Proc. of the NATO Advanced Study Institute, Erice 1975,
	eds. G. Velo and A.S.Wightman (Reidel, Dordrecht/Boston 1976,
	p.371); A.M. Polyakov, {\it Nucl. Phys.} {\bf 120} (1977)
	429; S. Samuel, {\it Phys. Rev. } {\bf D18} (1978)
	1916
\item{[11]} P. Minnhagen, {\it Phys. Rev. }{\bf B23} (1981)
	5745
\item{[12]} T.W.B. Kibble, {\it Phys. Rep.} {\bf 67}  (1980) 183;
	    J. Preskill and L. Krauss, {\it Nucl. Phys.}
	    {\bf B341} (1990) 50
  \item{[13]} G.E. Volovik and V.P. Mineev, {\it Zh. Eskp.
	Teor. Fiz. } {\bf 72} (1977) 2256  [{\it Sov. Phys. JETP } {\bf 45}
	(1977) 1186 ]
\end{description}
\end{document}